\documentclass{article}

\usepackage{graphicx}
\usepackage{amsmath}
\usepackage{wasysym}
\usepackage{caption}
\usepackage{subcaption}
\usepackage{comment}
\usepackage{breqn}
\usepackage{authblk}
\makeatletter
\renewcommand\@biblabel[1]{}
\makeatother

\newcommand{\be}{\begin{equation}}
\newcommand{\ee}{\end{equation}}
\newcommand{\nlm}{{\text{nlm}}}

\author[1]{Steve Markham\footnote{smarkham@caltech.edu}}
\author[2]{Daniele Durante}
\author[2]{Luciano Iess}
\author[1]{Dave Stevenson}
\affil[1]{California Institute of Technology, GPS}
\affil[2]{Sapienza University, Rome}
\title{Possible evidence of p-modes in Cassini measurements of Saturn's gravity field}

\begin{document}
\maketitle

\begin{abstract}
We analyze the range rate residual data from Cassini's gravity experiment that cannot be explained with a static, zonally symmetric gravity field. 
In this paper we reproduce the data using a simple forward model of gravity perturbations from normal modes. 
To do this, we stack data from multiple flybys to improve sensitivity. 
We find a partially degenerate set of normal mode energy spectra which successfully reproduce the unknown gravity signal from Cassini's flybys. 
Although there is no unique solution, we find that the models most likely to fit the data are dominated by gravitational contributions from p-modes between 500-700$\mu$Hz. 
Because f-modes at lower frequencies have stronger gravity signals for a given amplitude, this result would suggest strong frequency dependence in normal mode excitation on Saturn. 
We predict peak amplitudes for p-modes on the order of several kilometers, at least an order of magnitude larger than the peak amplitudes inferred by Earth-based observations of Jupiter. 
The large p-mode amplitudes we predict on Saturn, if they are indeed present and steady state, would imply weak damping with a lower bound of $Q>10^7$ for these modes, consistent with theoretical predictions. 
\end{abstract}

\section{Introduction}
Gravity field measurements allow us to probe the interior structure of a planet by measuring its deviation from spherical symmetry. 
For giant planets, the planet's response to its own rotation breaks its spherical symmetry. 
The deviation away from spherical symmetry depends on the planet's internal density distribution (\cite{dave-book}). 
Therefore a detailed mapping of a planet's gravity field can corroborate or refute interior models. 
Saturn's non-spherical gravity field was first inferred from spacecraft tracking data of Pioneer 11 (\cite{null+1981}, \cite{hubbard+1980}), and were later improved using Voyager data (\cite{campbell-anderson1989}). 
The arrival of the Cassini spacecraft in the Saturnian system yielded more accurate determination of the gravity field of the gas giant by first looking at the orbits of its satellites. 
Now the Grand Finale of the Cassini mission has produced exquisite gravity field data for Saturn, providing the first concrete constraints for Saturn's ring mass, zonal wind depths, and evidence for internal differential rotation by offering gravity field measurements up to $J_{12}$ (\cite{iess+2019}, \cite{galanti+2019}). 
But behind these spectacular new findings lurks a dark side: a small component of Saturn's gravity field which cannot be explained with the canonical static, zonally symmetric gravity field expected of gas giants. \\

Cassini's radioscience experiment is carried out by measuring the Doppler shift of a microwave signal in a two-way configuration: the signal is sent from a ground station to the spacecraft, which retransmit it back to the station preserving phase coherency. The Doppler shift is, to first order, proportional to the relative velocity of the spacecraft with respect to the station. These measurements are compared with predictions based on dynamical and observation models to obtain data residuals. The data we used in this study are two-way Doppler residuals, converted in a radial velocity time series, obtained by removing the effect of empirical acceleration from the reference solution given by Iess, et al., 2019.\\

This additional and unknown source of gravity can be fit with a variety of models. 
A static tesseral gravity field is possible, but there is no convincing low order fit (\cite{iess+2019}). 
A low order tesseral field does not provide a predictive solution with the available data, and also depends on the assumed rotation rate of Saturn. 
That is, a given gravity harmonic solution for a subset of flybys will not accurately predict the next flyby and requires additional harmonic terms. 
The nominal method which was employed for the published gravity harmonic results was an agnostic ``empirical acceleration'' model which, due to the unknown origin of the source of the additional gravity, included random acceleration vectors which changed on a ten minute timescale. 
In this context ``random'' means that each acceleration vector is allowed to have any direction with an \textit{a priori} amplitude of $\pm 4 \times 10^{-10}$km/s (\cite{iess+2019}). 
They could be correlated (non-random) even when the process used to create them allows for randomness. 
This timescale between changing acceleration vectors was determined empirically as the longest timescale which can successfully reduce range rate residuals to the noise level. \\

A time dependent signal does not necessarily require normal modes. 
For example, there may be a time dependent or non-symmetric signature from large scale convection (\cite{kong+2016}). 
Additionally, Saturn's envelope is differentially rotating (\cite{galanti+2019}, \cite{chachan-stevenson2019}). 
If mass anomalies were embedded at different depths or latitudes, then a spacecraft could encounter measurably different quasi-static tesseral gravity fields during each flyby (\cite{iess+2019}). 
However, differentially rotating tesseral structure in Saturn's gravity field has been shown to produce structures in the rings (\cite{elmoutamid+2017}), and the magnitude of the potential perturbation inferred from observation is orders of magnitude too small to explain the anomalous signal. 
Because of Saturn's expected internal differential rotation rate (about 5\%) (\cite{galanti+2019}, \cite{chachan-stevenson2019}), it is unlikely that such structure could measurably affect the spacecraft trajectory without showing clear structure from resonances in the rings. \\

This work will specifically explore the hypothesis that Saturn's residual gravity is a consequence of normal mode oscillations. 
It has already been demonstrated that normal modes are capable of eliminating the range-rate residuals to the noise level (\cite{iess+2019}). 
This has been done by computationally optimizing for individual mode amplitudes using a large number of free parameters. 
One possible solution involves only zonal f-modes. 
This solution, however, is affected by model assumptions such as maximum modeled spherical degree, whether to permit p-modes or g-modes, whether to permit non-zonal normal modes, etc. 
These uncertainties occur because, when optimizing with a large number of free parameters, there is a risk of over-fitting the data using too complex of a model. 
These issues are not important in the context of constraining Saturn's zonal gravity harmonics and ring mass because the uncertainty can simply be absorbed in the error ellipses for these values. 
However in this work we revisit the residuals data with a different purpose: to try to extract a preferred normal mode spectrum which is predictive for further flybys, robust to changes in model assumptions, and as simple as possible to capture the qualitative behavior of the spectrum without over-fitting the data. 
Bearing this in mind, although we find a statistical preference using our simple model for signals dominated by low-order p-modes, readers should remember that our findings are not conclusive proof of such a spectrum on Saturn. 
\\

Our investigation has at least two important applications: first, normal modes are themselves a promising method by which to probe the interior structure of giant planets, and this analysis provides some evidence of their power spectrum. 
Second, any gravitational signal from normal modes above the noise level contaminates spacecraft tracking data and may be aliased into the static model. 
As we will see, the behavior of the modes are partially degenerate and the solution is non-unique. 
However, the solutions are clustered in parameter space and predict a high probability of reproducing the observed unexplained gravity signal. 
The most successful models 
indicate the signal is likely to be dominated by p-modes between 500 and 700$\mu$Hz (see Figure~\ref{degeneracy}). \\


In the second section, we outline some fundamentals of giant planet seismology, spacecraft tracking, and our forward model. 
In the third section we discuss our data reduction method including a novel data stacking technique, as well as error sources, and a fitting procedure. 
In the fourth section we present the results of our analysis, finding a simple two parameter model which has a high probability of producing a good fit to the spacecraft signal. 
In the final section we discuss the implications of our findings.

\section{The forward model}\label{forward}
In order to accurately model seismic effects on Cassini's gravity signal, we must determine the mode's eigenfrequencies, and the scaling relationship between a mode's displacement amplitude and its effect on Saturn's gravity field. 
These issues are addressed in Section~\ref{background}. 
Next we must model how a given gravity potential perturbations affects the spacecraft tracking signal, which is done in Section~\ref{integ}. 
Next we need an agnostic parametric model for modal energy spectrum, discussed in Section~\ref{spectral}. 
Finally we account for the intrinsic stochasticity of the problem; a given mode cannot be modeled deterministically, because we have no way of knowing what the temporal phase of each mode was when Cassini was at periapse. 
This is partially circumvented with our stacking technique, discussed in Section~\ref{stacking} with further technical information in the Appendix. 

\subsection{Background}
\label{background}
In this paper we approximate Saturn as an adiabatic, spherical, uniformly rotating planet. 
We neglect rotation to compute the eigenfunctions and potential perturbations, but account for rotation when considering Coriolis force frequency splitting and the rotating gravity potential encountered in an inertial frame. 
In this case, giant planet oscillations can be decomposed into a discrete set of orthogonal normal modes with quantum numbers $(n,l,m)$. 
$n$ corresponds to the number of radial nodes in the displacement eigenfunction, $l$ to the spherical harmonic degree, and $m=[-l..l]$ to the azimuthal degree. 
Each mode has a unique displacement eigenfunction 
\begin{equation}
\xi_{\text{nlm}}(\mathbf{r}) = \left( \xi_{r\text{,nlm}}(r)\hat{\mathbf{r}} + \xi_{h\text{,nlm}}(r)\hat{\theta}\frac{\partial}{\partial \theta} + \frac{\xi_{h\text{,nlm}}(r)}{\sin\theta}\hat{\phi} \frac{\partial}{\partial \phi} \right) Y_{lm}(\theta,\phi)
\end{equation}
and a characteristic eigenfrequency $\omega_{\text{nlm}}$ so that the total displacement as a function of time is $\xi_{\text{nlm}}\cos(\omega_{\text{nlm}} t + \alpha_{\text{nlm}})$. 
Because $\omega_{\text{nlm}}$ is not precisely determined, the phase $\alpha_{\text{nlm}}$ cannot be coherently specified between flybys and is assumed random for each mode for each flyby. 
$\xi_r$ and $\xi_h$ correspond to the radial and horizontal eigenfunctions respectively, which together specify the fluid displacement at any point within the planetary sphere. 
For our purposes the eigenfunctions were obtained using GYRE stellar oscillation code (\cite{gyre}), with an n=1 nonrotating polytrope model for Saturn's interior. 
Our goal here is independent of accurate interior modeling; we are interested in the relative gravity signal between modes and their order of magnitude, which is not strongly sensitive to small changes in the interior model. 
However, using an adiabatic interior model precludes g-modes, so we account for contributions from g-modes separately. \\

\begin{figure}
\centering
\includegraphics[scale=.8]{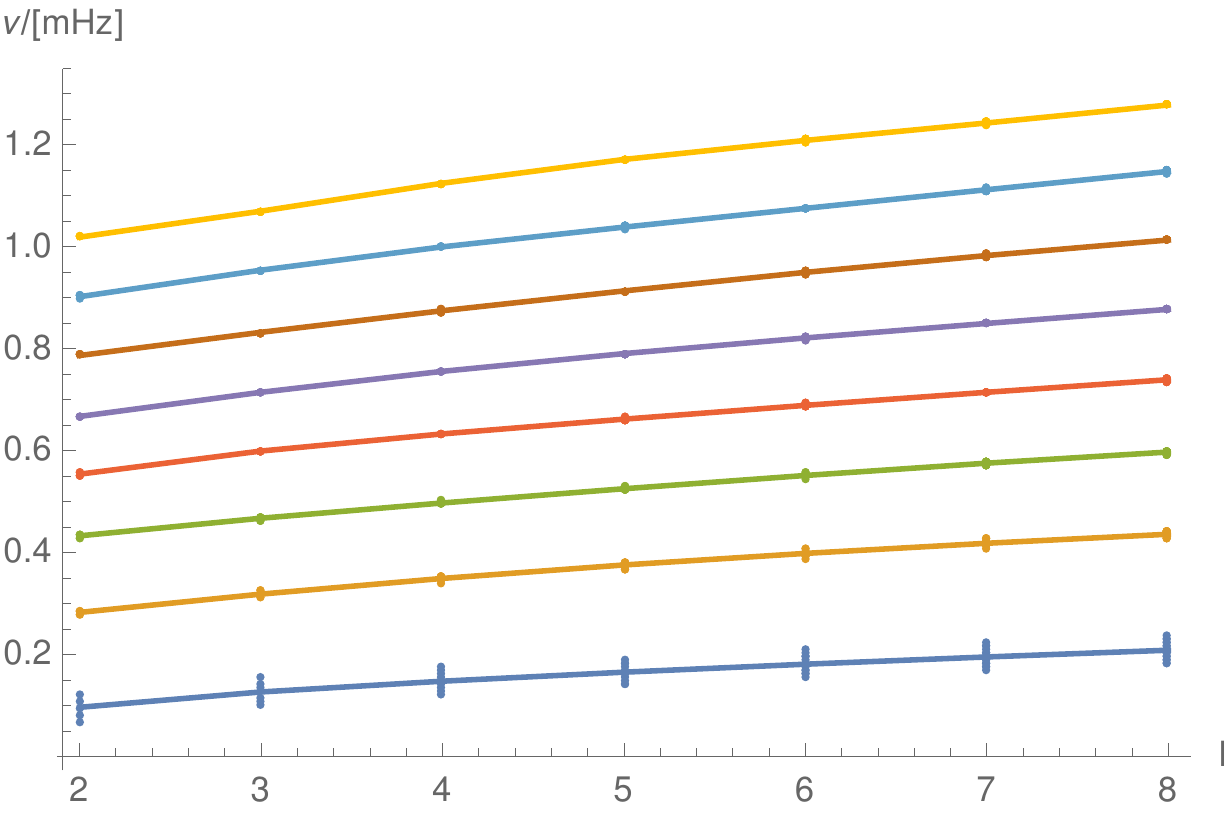}
\caption{Eigenfrequencies including splitting due to the Coriolis force in the rotating frame. Each curve corresponds to rising radial order $n$ for $m=0$ modes, with $l$ rising along the x-axis.  
Each eigenfrequency $\omega_{\text{nlm}}$ is shown as a point, with $m\neq 0$ modes deviating from the $m=0$ curve. 
This frequency splitting effect is most important for f-modes. 
}
\label{freqs}
\end{figure}

Because Saturn's interior structure is not precisely determined, 
we performed our full analysis on a variety of interior model assumptions 
to demonstrate that the results are not sensitive to small errors in modal eigenfrequencies. 
We tested eigenfrequencies produced by this same polytrope model generated with GYRE (\cite{gyre}), as well as a sampling computed using a more sophisticated Saturn interior model (\cite{gudkova-zharkov2006}). 
The nominal model uses the eigenfrequencies from Sa8. 
In addition, we accounted for mode-splitting due to Coriolis forces (\cite{christensen-dalsgaard}). 
These split according to 
\begin{equation}
\delta \omega_{\text{nlm}} = m \beta_{\text{nl}} \Omega
\end{equation}
where $\Omega$ is Saturn's spin rate and
\begin{equation}
\beta_{\text{nl}}\equiv \frac{\int_0^R(\xi_r^2+l(l+1)\xi_h^2-2\xi_r\xi_h-\xi_h^2)r^2\rho dr}{\int_0^R (\xi_r^2+l(l+1)\xi_h^2)r^2\rho dr}
\end{equation}
The nominal frequencies for this paper and plotted in Figure~\ref{freqs}.
For computational reasons, we consider a finite subset of modes in our model. 
In the nominal model we consider f-modes and p-modes up to $l=8$, $n=7$.
We found equivalent results when using instead $l=10$, $n=5$ as bounds on parameter space. 
We also specially tested f-modes only up to $l=20$.
We do not expect g-modes to dominate the signal for at least two reasons: 
first, because the stable layer where they resonate is so deep, its effect on the gravity field would be very weak unless its amplitude were extremely large. 
Second, we do not expect its amplitude to be extremely large, because its eigenfunction is evanescent near the surface where mode excitation is expected to be most efficient.
Nevertheless for the sake of completeness we tested g-modes using published eigenfrequencies (\cite{gudkova-zharkov2006}). 
 \\

\begin{figure}
\centering
\includegraphics[scale=.8]{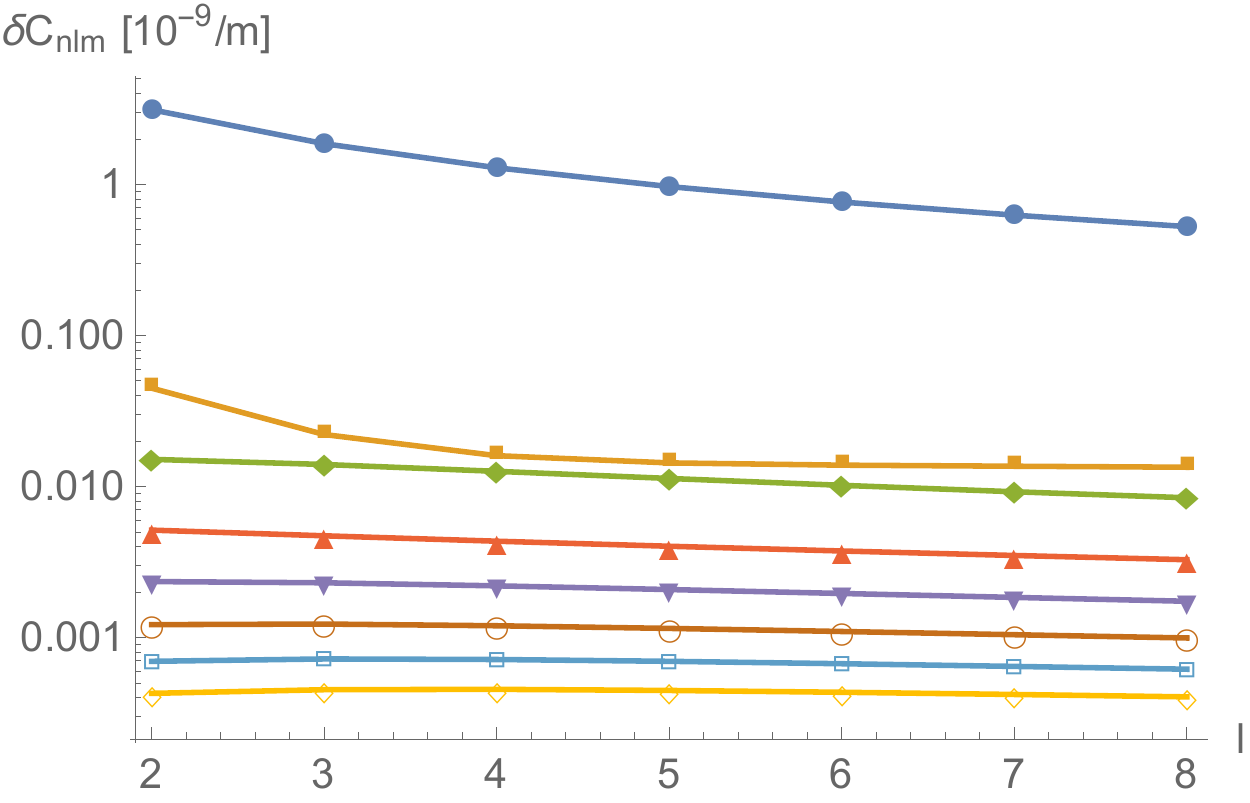}
\caption{Gravity harmonic coefficient perturbations for various modes per meter of surface displacement amplitude. To (top) blue curve represents f-modes which has the most prominent gravitational signature for a given surface amplitude, while p-modes (below) need larger amplitudes to be detected. }
\label{dpotss}
\end{figure}

After choosing eigenfrequencies, we compute the scaling between displacement eigenfunctions and gravity potential perturbations. 
The gravity field perturbation associated with displacement eigenfunction $\xi$ can be obtained by integrating over the material sphere and accounting for fluid point displacements according to 
\begin{equation}
\delta \Phi = G \int \frac{\rho(\mathbf{r}')}{|\mathbf{r}-(\mathbf{r}'+\xi(\mathbf{r}'))|}d^3r'-\frac{G M}{r}
\end{equation}
for linear perturbations this is equivalent to the Eulerian density perturbation from continuity $\delta \rho = \nabla \cdot (\rho \xi)$ so that 
\begin{equation}
\delta \Phi = G \int \frac{\delta \rho(\mathbf{r}')}{|\mathbf{r}-\mathbf{r}'|}d^3r'
\end{equation}
Decomposing this potential perturbation into 
\begin{dmath}
\delta \Phi = \frac{G M}{r} 
\sum_{n=0}^{\infty} \left( -\sum_{l=2}^{\infty} \sum_{m=-l}^l \left(\frac{R}{r}\right)^l P_l^m(\cos\theta) [\delta C_{\text{nlm}} \cos m \phi + \delta S_{\text{nlm}} \sin m \phi] \right)
\end{dmath}
one can show that the gravity harmonic coefficient perturbation associated with the normal mode is (\cite{dave-book}, \cite{marley-porco1993})
\begin{equation}
\delta C_{\text{nlm}}(t)= \frac{1}{M R^{l}} \frac{2l+1}{4\pi} \cos(\omega_{\text{nlm}}t) \int_0^R r^{l+2} \delta \rho_{\text{nlm,}r}(r) dr
\end{equation}
where $\delta \rho_{\text{nlm}}(\mathbf{r}) \equiv \delta \rho_{\text{nlm,}r}(r) P_l^m(\theta)\cos(m \phi)$ normalized such that the mode surface displacement is 1cm at the planet surface. 
With appropriate choice of coordinates, $\delta S_{\text{nlm}}\rightarrow 0$. 
This leads to Figure~\ref{dpotss} which illustrates why f-modes are a priori favored as sources of gravity perturbations. 
Higher order p-modes have nodes in their eigenfunction, leading to destructive interference of the gravitational signature. 
Therefore in order for p-modes to dominate the signal, they must have more than an order of magnitude larger energy than f-modes (two orders of magnitude larger amplitude). 
\\

A test particle outside of the planet on a prescribed trajectory $\mathbf{r}(t)$ encounters the potential perturbation
\begin{dmath}
\delta\Phi (t) = \frac{G M}{r} 
\sum_{\text{nlm}} a_{\text{nlm}} \delta C_{\text{nlm}} \left(\frac{R}{r(t)}\right)^{l+1} P_{lm}(\cos\theta(t))\\
\cos[m(\phi(t)-\Omega t-\phi_0)-\omega_{\text{nlm}}t-\alpha_{\text{nlm}}]
\label{mode-pot}
\end{dmath}
where $a_{\text{nlm}}$ is the maximum surface displacement of mode $(n,l,m)$ in cm, and $\Omega$ is Saturn's spin rate. 
Notice that this equation includes two random variables: $\phi_0$ the initial longitudinal orientation of the modes with respect to our coordinates, and $\alpha_{\text{nlm}}$ is the initial temporal phase of the mode. 
$\phi_0$ is shared between all modes, but is random for each flyby. 
$\alpha_{\text{nlm}}$ is a random variable for each mode and for each flyby. 
Although the phase difference between flybys can in principle be determined from the mode's eigenfrequency, in practice this is impossible. 
Eigenfrequencies depend on Saturn's interior structure and cannot be predicted with perfect precision. 
Because the time between encounters is much longer than the period of a mode, in practice the phase of each mode must be regarded as randomized for each flyby. 
This stochasticity introduces a complication for modeling the flyby--we do not know the initial phase and orientation of the modes when each flyby occurred. 
This issue will be addressed in Section~\ref{data-stacking}. 

\subsection{Numerical integration and model reproduction}
\label{integ}
The gravity experiments were conducted with an edge-on geometry from Earth's perspective for maximum signal to noise. 
The orbits were highly inclined and highly eccentric. 
The closest approach (C/A) of the spacecraft is about 5\% of Saturn's radius from its cloud tops, approaching and receding from the planet very quickly during ingress and egress. 
Because of this orbital geometry, we only expect a significant signal from the planet within about an hour of C/A. 
Therefore we use the spacecraft orbital elements two hours before C/A to compute the initial conditions. 
We then numerically integrate the equation of motion for Saturn plus the potential perturbation associated with normal modes, neglecting oblateness. 
Including the measured static zonal gravity (\cite{iess+2019}) affects the simulated normal mode range rate signal by less than 1\%. 
We integrate the equation of motion using Mathematica's built in integrator to generate a three dimensional velocity time series. 
We then subtract the Keplerian solution from the numerical solution to isolate the signal from the spectrum of modes we are modeling. 
Finally we project the three dimensional velocity vector onto the line of sight vector connecting Saturn to the Earth. 
We verify the accuracy of this method by reproducing the signature from static zonal J's (\cite{iess+2019}) using the method from this paper, finding good agreement. 
This method is fully general for any potential perturbation, and we will use it to inspect the behavior of normal modes. \\

We verified empirically the approximate linearity of combining the velocity perturbation from various sources. 
That is,
\begin{equation}
\delta \mathbf{v} (t) = \sum_q \mathbf{v}_q(t)
\label{velocity-superposition}
\end{equation}
within $<$0.1\% for the perturbation magnitudes in question. 
Strictly speaking this linearity does not hold absolutely; although gravity potential perturbations are exactly linear, a test particle encountering these perturbations may be perturbed from its trajectory--if this perturbation is sufficiently large the linearity breaks down. 
But for the small perturbation of interest, this non-linearity is not important. \\

It is important to discuss at this point a fundamental ambiguity in probing for normal modes from the spacecraft's perspective. 
The spacecraft is observing two sources of variation of the gravity signal: the intrinsic geometric variation, and the temporal variation. 
The geometric variation is the physical shape of the mode, which attenuates with distance and varies with the spacecraft's latitude and longitude relative to Saturn. 
As the spacecraft approaches and recedes from the planet, traveling from the north to south and west to east, even a static gravity perturbation would have a time dependent signal from the spacecraft's frame of reference. 
On the other hand, the potential perturbation itself varies with time. 
The convolution of these effects makes it difficult to have a simple intuition for Cassini's response to each mode. 
See the Appendix for further discussion. 

\subsection{Spectral model}
\label{spectral}
For the spectral model, we aim to be as agnostic as possible. 
We do not know with certainty by what mechanism seismic activity is excited on Saturn, although meteor impacts (\cite{wu-lithwick2019}) or exotic meteorological phenomena (\cite{markham-stevenson2018}) have been suggested. 
Therefore we only use simple parametric models which scale the energy of each mode as a function of parameters. 
We tested a variety of scaling relationships, including power law dependence in frequency, as well as power law dependence on quantum numbers $n$, $l$, and $\frac{l-|m|}{l}$. 
We also tested equipartition. 
None of these models provided convincing fits to the data. \\

The nominal model used is a Gaussian frequency dependent model, although here too we aimed to be as agnostic as possible. 
The Gaussian is convenient because it has the power to probe for a diverse variety of frequency dependencies by only varying two parameters. 
Using this assumption, the mode energy is a function of its eigenfrequency according to
\begin{equation}
E(\omega_{\text{nlm}}) = \frac{1}{\sqrt{2 \pi} \sigma} \exp\left[-\frac{(\omega_{\text{nlm}}-\omega_0)^2}{2 \sigma^2} \right]
\label{spec-eq}
\end{equation}

We tested extreme parameters, varying the peak of the Gaussian between 0 and 5mHz, well above the acoustic cutoff frequency. 
We also tested widths between an extremely narrow distribution of 10$\mu$Hz and an extremely wide distribution of 5mHz (see Figure~\ref{probs}). 
By varying the parameters so widely, we can capture a wide variety of possible frequency dependent behavior.
A Gaussian with a far away peak behaves approximately like an exponential relationship. 
A Gaussian with an extremely narrow peak behaves approximately like a $\delta$-function, and one with an extremely wide peak approximates equipartition. 
As discussed in Section~\ref{results}, we find narrowly peaked distributions centered on low order p-modes to be the most likely to fit the data, although there is considerable degeneracy within that region of parameter space. 
We settled on the Gaussian dependence on frequency after trying a variety of parametric models because it provided the best fit to the data, and is flexible in qualitatively approximating many diverse behaviors. 
We do not claim that the real power spectrum behaves in exactly this way.

This frequency dependence on energy has a straightforward connection to excitation and dissipation efficiency, if the energy spectrum is in steady state. 
In this case, 
\begin{equation}
E(\omega) = \frac{\dot{E}_{\text{in}}(\omega) Q(\omega)}{\omega}
\end{equation}
where $Q(\omega)$ is the frequency dependent quality factor and $\dot{E}_{\text{in}}(\omega)$ is the frequency dependent excitation rate. 
The mode energy is $E_{\text{nlm}}=a_{\text{nlm}}^2 \omega_{\text{nlm}}^2 M_{\text{nlm}}$ where $M_{\text{nlm}}$ is the modal inertia uniquely defined for each normal mode according to $M_{\text{nlm}}=\int \rho |\xi_{\text{nlm}}|^2 d^3r$, where $\xi_{\text{nlm}}$ is the mode eigenfunction normalized such that the surface displacement in 1cm. 
Therefore the energy scales as the square of the amplitude. 

\section{Data stacking}\label{data-stacking}
\label{stacking}
There is a fundamental ambiguity when modeling normal modes that does not exist for a static gravity field: the phase of the mode in question. 
If we only had one flyby, breaking this ambiguity would be hopeless; since we have multiple, we can do better. 
By combining multiple flybys, we can average out the effect of initial phases. 
This can be done perfectly if there are a large number of identical flybys. 
Indeed, one can show (see Appendix) for a particle on a prescribed trajectory $\mathbf{r}(t)$ encountering a potential perturbation of the form of Equation~\ref{mode-pot} where $\phi_0$ and $\alpha_{\text{nlm}}$ are random variables for each flyby, that the summed squared potential obeys the asymptotic relationship 
\begin{equation}
\sum_i^N \delta \Phi_i^2(t) \sim \frac{N}{2} \sum_{\text{nlm}} A_{\text{nlm}}^2 f_{\text{nlm}}^2(t)
\end{equation}
where $A_{\text{nlm}} \equiv a_{\text{nlm}}\delta C_{\text{nlm}}$, and $f_{\text{nlm}}(t)$ is a deterministic function of time independent of $\phi_0$ and $\alpha_{\text{nlm}}$. 
This approximation is valid at large $N$. 
It is possible to derive a similar expression for acceleration perturbations, simply the gradient of the potential perturbations, and for velocity perturbations (see Appendix). 
In fact these derivations depend on assuming a prescribed trajectory, but in reality the potential perturbations perturb the trajectory itself. 
We verify empirically that for a sufficiently large number of flybys
\begin{equation}
\sum_i^N \delta v_i^2 = \frac{N}{2} \sum_{\text{nlm}} a_{\text{nlm}}^2 \sum_i^N \delta v_{\text{nlm,i}}^2
\label{velocity-stack}
\end{equation}
where $\delta v_{\text{nlm,i}}^2$ is a randomly generated squared time series of velocity perturbation associated with the spacecraft encountering the potential perturbation due to a 1cm displacement amplitude mode with quantum numbers $(n,l,m)$. \\

We note that the stochastic behavior of the modes approaches deterministic behavior when summing over a large number of flybys to demonstrate why such an exercise is useful: it reduces the stochastic component of the signal and amplifies the deterministic component. 
In the real experiment, however, there were only five flybys, which is not large enough to simply stack the data and compare it against the asymptotic average. 
Therefore, we ran a Monte-Carlo simulation, leaving the initial phase as a free random variable, and combined the signal from five randomly selected flybys with an input spectrum scaled according to Section~\ref{spectral}. 
In this case, we use the relationship from Equation~\ref{velocity-superposition} to obtain
\begin{equation}
\sum_i^N \delta v_i^2 = \sum_i^N \left( \sum_{\text{nlm}} a_{\text{nlm}} \delta v_{\text{nlm}} \right)^2
\label{velocity-exact}
\end{equation}
this expression is equivalent to Equation~\ref{velocity-stack} in the limit of large $N$, but for finite $N$ has stochastic components which should be accounted. \\

With $N=5$ we can eliminate significant ambiguity. 
The raw data is shown in the red scattered points in Figure~\ref{flyby-fits} for each flyby (the black curves are model fits to the data).
The raw data were obtained by subtracting the observed spacecraft signal from a model excluding stochastic acceleration (\cite{iess+2019}). 
We take these points and bin them into 150 second windows so that most points in time will have contributions from all flybys (see Section~\ref{errs} for why this is important).
We then average the square value of the corresponding data point across the five flybys to obtain an average value. 
After accounting for various quantifiable sources of error, we produce Figure~\ref{example} that shows the stacked data with error bars in red, with a black curve as a good fit forward model. \\

\subsection{Error sources}\label{errs}
To average this data, we must propagate the errors from the input data, and account for additional errors from the stacking process. 
We have identified three quantifiable sources of error in the tracking system, which we use for the error bars. 
The first source of error is the intrinsic noise in the system (\cite{iess+2019}).
This source of error affects all data points. \\

The second source of error is the fact that part of the ``real'' non-static, non-zonal gravitational signal may have been aliased into the uncertainty about the static zonal gravity harmonic coefficient J's (\cite{iess+2019}). 
To understand how this impacts the data, we ran a Monte Carlo simulation systematically adding the gravitational signal $\delta J_l$ for each zonal gravity harmonic $l$, and running that modified data through our stacking pipeline. 
We modeled each $\delta J_l$ as a statistically independent normally distributed random variable using the published 1$\sigma$ formal uncertainty (\cite{iess+2019}) (although the total value for different $J_l$'s are correlated, small deviations $\delta J_l$ can be approximately independent). 
We found the impact of this effect by taking the standard deviation of the stacked data for 1000 such simulations and used those values as an additional independent source of error to add in quadrature with instrumental noise. 
This source of error is most important near closest approach. \\

The third source of error is only applicable to a subset of points, but is the most important source of error for those points. 
Because we will be comparing this data to simulations without gaps or sampling issues, we need to account for the fact that some data points do not average all five flybys. 
This occurs because the time window in which the spacecraft is blocked by Saturn's rings is slightly different for each flyby, and because some of the data sets end before others. 
When we only average a subset of data points together, there will be a systematic offset from the otherwise smooth behavior of the average. 
We quantify this offset by taking samples of points that have data from all five flybys, then calculate the average systematic offset caused by using only a subset of those data points. 
We use this average value as an additional source of error, which is simply a function of the number of data points averaged. 
If there are five data points, this source of error is zero.
Note that this error is systematic, so a series of points all missing one data set will not be randomly scattered around the main curve but will be systematically offset from it. 
Accounting for these three quantified sources of error produces the red points and error bars in Figure~\ref{example}, which is the time series data set we will attempt to reproduce (with an example black curve model fit). \\

There are additional sources of error that are likely to prevent us from getting a perfect fit to the data. 
First, we do not know the actual eigenfrequencies of the modes. 
We attempted multiple assumptions for the frequencies to verify that our conclusion is not affected by different choices. 
In fact, the signature from a flyby is a slowly varying function of mode frequency, so expected errors (less than 10\% in frequency) should not affect the general, qualitative behavior of the flyby signature. 
Nevertheless, errors in frequency yield systematic modeling errors, small temporal offsets for small errors and slowly varying qualitative behavior for larger errors, such that the fit will not be perfect.
This impacts the goodness of fit. 
An additional source of possible modeling error is the simplicity of our assumptions (a smoothly varying amplitude spectrum). 
For example, the excitation mechanism may be partly stochastic in nature (\cite{goldreich+1994}, \cite{markham-stevenson2018}), and the real frequency dependence may be more complicated or jittery than a simple Gaussian. 
Accounting for this possibility, however, would violate the purpose of this investigation: to keep the number of parameters small, and the spectral model simple. 
Another cause of error we have not formally accounted is the difference in geometry between the flybys. 
To first order, the orbit is similar and the Saturn-Earth orientation is nearby during each flyby. 
But the subtle differences in geometry means we should not expect the assumption of fixed geometry and identical orbit initial conditions to reproduce the data exactly. 
Nevertheless, this is a necessary assumption in order to use the stacking method to amplify the deterministic component of the signal. 
We note these sources of error not to rigorously quantify their effect, but to justify our relatively lax error tolerance for goodness of fit; the upshot here is that we are trying to evaluate the probability of reproducing the general qualitative behavior of the signature for a given power spectrum, not to provide a single exact reconstruction of the gravity field Cassini encountered (doing so would be impossible with the available data anyway).

\section{Analysis and results}\label{results}
Now that we have added error bars to account for the straightforwardly quantifiable sources of error, we can attempt to fit them. 
We do this with a reduced $\chi^2$ test according to 
\be
\chi^2 = \frac{1}{D-f} \sum_i^D \frac{(x_i - m_i)^2}{e_i^2}
\ee
where $D$ is the number of data points, $f$ is the number of degrees of freedom in the model (three in our case: the two Gaussian parameters and the scaling coefficient), $x_i$ is data point $i$, $m_i$ is its corresponding modeled value, and $e_i$ is the error. 
Choosing an appropriate model is subtle. 
One choice is to use the asymptotic average of an infinite number of flybys for a given spectral model. 
As demonstrated in Section~\ref{data-stacking}, we expect the data to converge toward this average. 
But given the finite number of flybys, there will be variation from this asymptotic mean. 
Therefore in order to evaluate the likelihood of a given model, we conducted 2,000 tests of five simulated flybys for each modeled spectrum. \\

\begin{figure}
\centering
\includegraphics[scale=.8
]{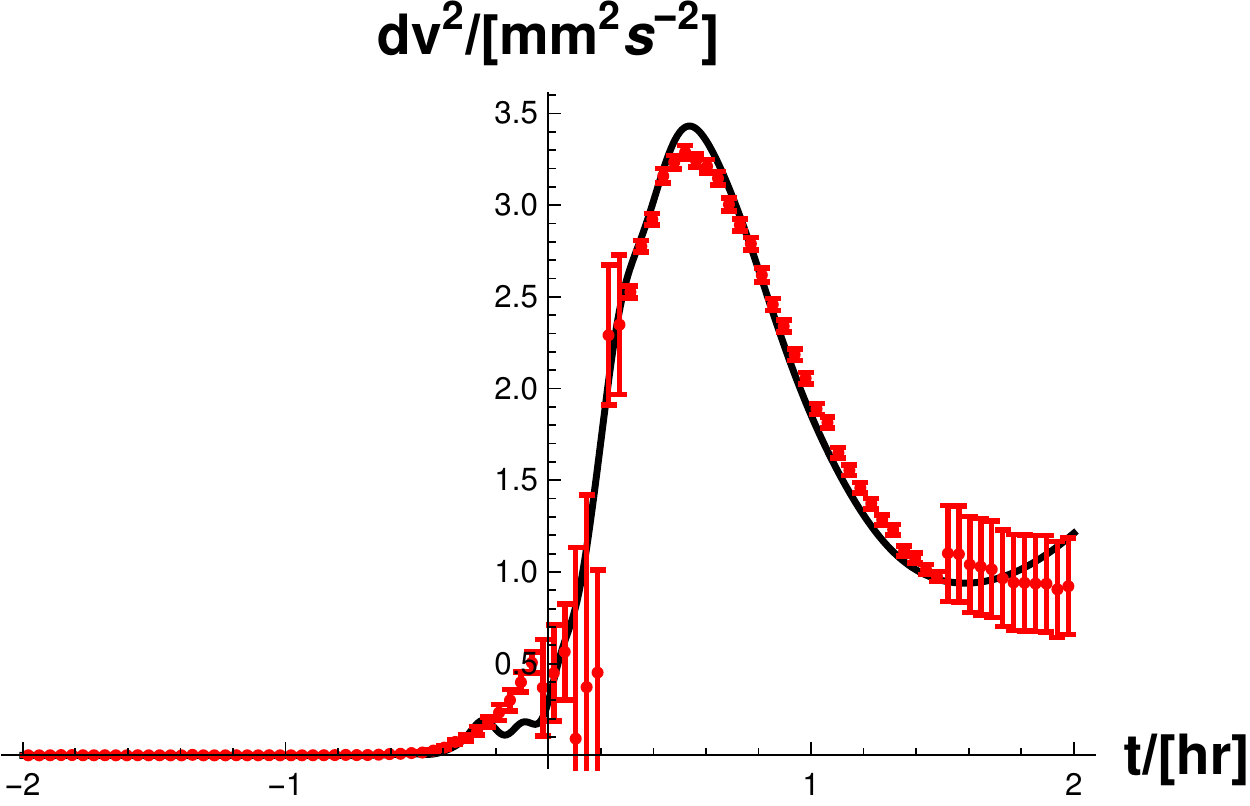}
\caption{An example fit to the stacked data. 
The data is represented with error bars according to Section~\ref{errs}, with the black curve corresponding to an example energy spectrum with $\omega_0=600\mu$Hz and $\sigma=40\mu$Hz.}
\label{example}
\end{figure}

To produce our forward model, we ran $10^4$ simulated signals from individual modes, for the subset of considered modes (recall our results are not sensitive to the specific choices of considered modes or computed eigenfrequencies. For more discussion see Section~\ref{forward}). 
After simulating a large number of range rate signals for each individual mode, we chose five to combine their squared signal using Equation~\ref{velocity-exact}, 
where $a_{\text{nlm}}$ for a given model is computed according to Equation~\ref{spec-eq} with $E_\nlm=a_\nlm^2 \omega_\nlm^2 M_\nlm$. 
This is the forward model we use to try to fit Figure~\ref{example}. 
Our tolerance threshold for goodness of fit is $\chi^2=50$. 
This is a large value, but we consider it sufficient to qualitatively reproduce the essential shape of the data (for further discussion as to why this is appropriate, see Section~\ref{errs}). 
Choosing a different threshold does not significantly affect the results, but reduces probability of fitting within the tolerance threshold for all models. 
We then use the large number of experiments to assign a probability of reproducing the data within tolerance for a given input spectrum. 
The results show a degenerate set of distributions which can reproduce the signal. 
We found a strongly favored region of parameter space after coarse sampling using extreme parameters, then followed up with a finer sampling in that region. \\

\begin{figure}
\begin{subfigure}{.5\textwidth}
  \centering
  \includegraphics[width=\linewidth]{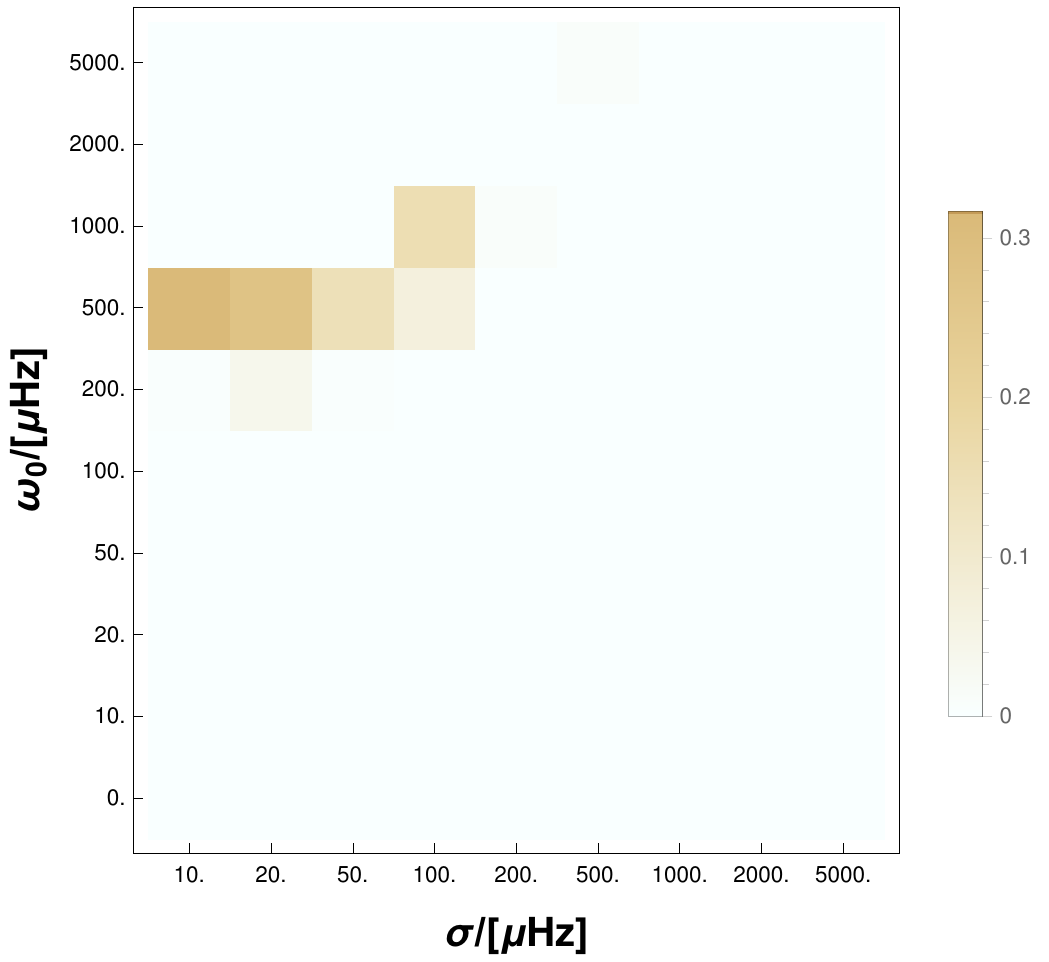}
  \caption{Coarse-grained grid search}
  \label{fig:sfig1}
\end{subfigure}%
\begin{subfigure}{.5\textwidth}
  \centering
  \includegraphics[width=\linewidth]{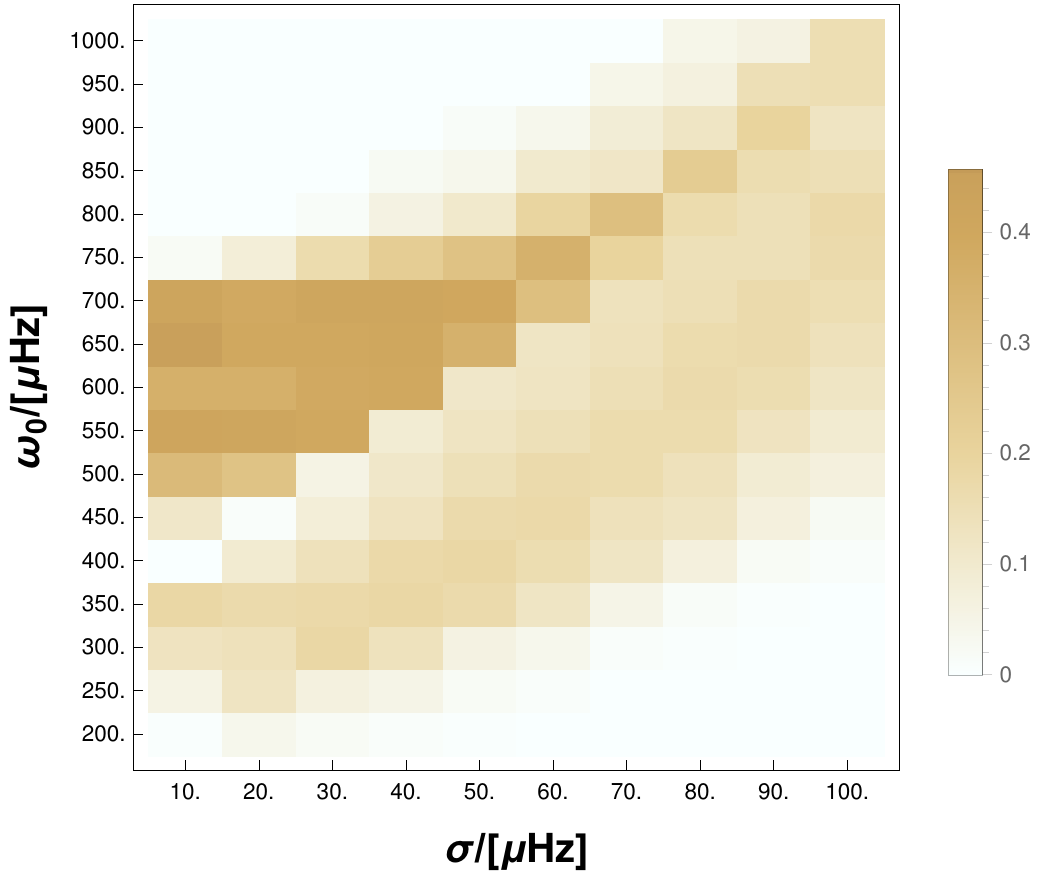}
  \caption{Fine-grained grid search}
  \label{fig:sfig2}
\end{subfigure}
\caption{Probability maps for different input parameters. Darker colors correspond to models which have a higher probability of satisfactorily fitting the data. (a) Left shows a coarse grained plot, which searches a wide range of parameter space including models which approximate exponential behavior, delta-functions, or white noise, indicating a preferred region of parameter space. (b) Right shows a finer grained sampling in this region, illustrating the degeneracy within that region.}
\label{probs}
\end{figure}

\begin{figure}
\centering
\includegraphics[scale=1]{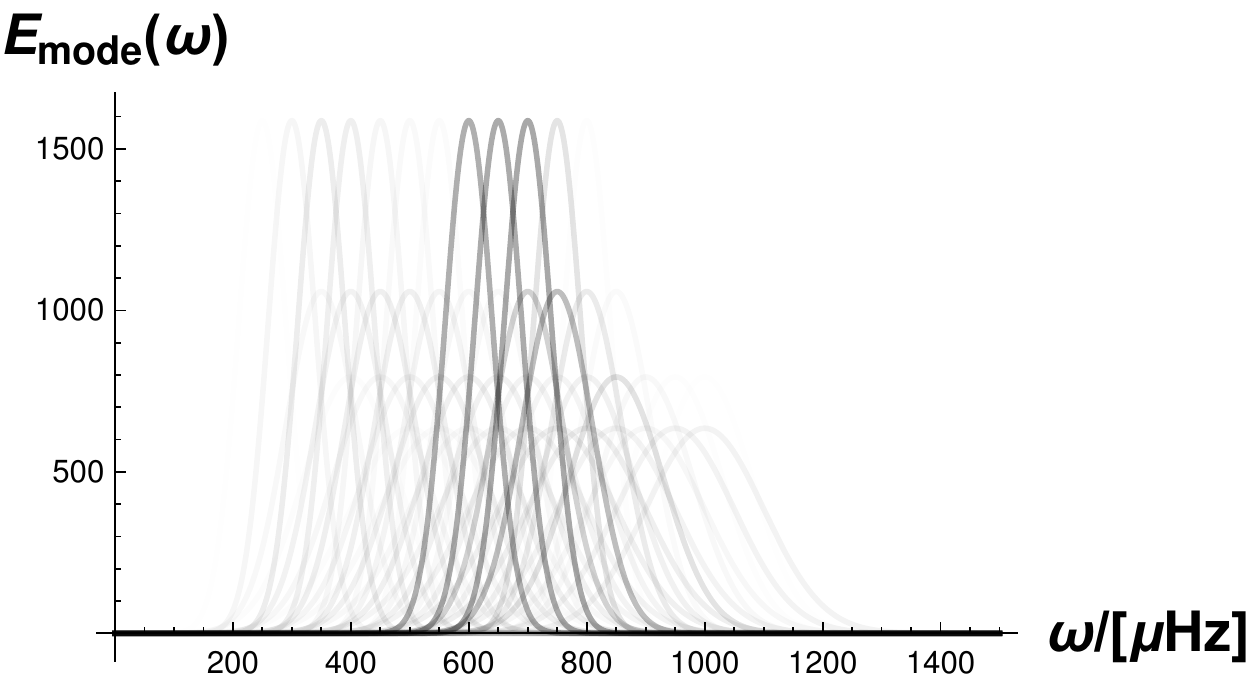}
\caption{Plotting different allowable solutions, with the darkness of the curve corresponding to the probability that, if that spectrum is correct, we would observe the data within our tolerance level. The curves are normalized in the plot such that their integrated value is unity (using Hz rather than $\mu$Hz as the ordinate).}
\label{degeneracy}
\end{figure}

The probability plots are show in Figure~\ref{probs}, which illustrates a clearly preferred region of parameter space, but degeneracy within that region. 
Each grid cell of the Figure~\ref{probs} represents a particular spectral model. 
The shading indicates the probability of reproducing the data within our tolerance threshold, if that spectrum were Saturn's normal mode spectrum. 
The degeneracy can be partly understood by considering the contributions to the gravity signal a moving spacecraft encounters. 
The degeneracy can be understood in two ways. 
First, as shown in Figure~\ref{degeneracy}, there is a great deal of overlap between favored models. 
Second, because the spacecraft is moving through space, a static field would have a time dependent signature. 
Because normal modes oscillate in time, there is another source of time dependence which would be experienced even by a stationary test particle. 
The synthesis of these two contributions allows gravity perturbations with different properties to produce a similar signal along the Saturn-Earth line of sight axis from the spacecraft's frame of reference. 
We elaborate on this second degeneracy source in Section~\ref{integ} and in the Appendix. \\

Although we cannot identify a single conclusive power spectrum, we can exclude a wide variety of simple spectra, and find the highest probability models favor a relatively narrowly peaked distribution. 
The location of the peak is also constrained, with the most likely models having a peak between 500 and 700 $\mu$Hz.
\\

Although the exact width and frequency peak cannot be precisely determined, we can exclude a wide variety of models as implausible, and note a clear clustering of models which have a high probability of reproducing the observed signal. 
Low order f-modes lie generally below 200$\mu$Hz (see Figure~\ref{freqs}). 
No good-fit models favor significant contributions from f-modes. 
We demonstrate an example of what we consider to be a ``plausible'' fit near the cutoff threshold in Figure~\ref{example}. 
This particular run has $\chi^2=4$, among the better fits we were able to obtain. \\

We also assessed the frequency content of the residual data. 
This proved to be less diagnostic than fitting the time series. 
This is perhaps unsurprising, because in order to fit the frequency content, you only need a model that varies on the correct timescales, driven largely by the geometric $\left(r/R\right)^l$ effect. 
By contrast, in order to fit the time series data, you have to match much more specific behavior. 
Looking at Figure~\ref{example}, the time series model must fit several specific phenomena. 
From left to right on the figure, the best models the steepness of the ``ramp up'' before C/A, the timing of the peak after C/A, the width of the main curve, the timing of the plateau/turnover, and other features. 
We find that the plausible time series fits are also compatible with the data's Fourier transform. 
However, the Fourier transform is much more degenerate and possible to fit with a wide variety of models, and it is difficult to obtain any new information. \\

To verify that our results were not excessively biased by our assumptions, we ran a variety of tests and alternatives, in addition to trying various parametric models as described in Section~\ref{spectral}. 
We also explored the possibility that the signal may be dominated by a single mode, by testing that hypothesis against each mode in our sample. 
This possibility seems plausible based on our results given the narrowness of the peak in many best fit cases. 
We found some modes within the preferred region of parameter space had a finite probability of reproducing the data, but the probability was lower than our preferred spectral models. 
Consistent with our spectral method, f-modes were not favored. 
No f-mode had a probability higher than 2\% of producing the observed signal. \\

We separately tested all f-modes up to spherical order 20 for completeness, because of the \textit{a priori} expectation that they should be the most gravitationally important modes, and because some of their frequencies overlap with the degenerate region of parameter space which can some probability of providing a tolerable fit to the data. 
High degree f-modes are discussed in more detail in the Appendix. 
Even allowing for higher order f-modes, we did not find any simple combinations that satisfactorily reproduce the data. 
Perhaps f-modes are inefficiently excited for reasons beyond their frequency; for example, some excitation models depend on compressibility (\cite{goldreich+1994}), and in the sun mode power declines for increasing $l$ even at fixed eigenfrequency. \\

We also tested g-modes. 
Although our interior model assumptions did not produce g-modes, we tested them specially using published eigenfrequencies (\cite{gudkova-zharkov2006}), and separately testing asymptotic approximations for their eigenfrequencies (\cite{tassoul1980}). 
Without the eigenfunctions, we tested the spectral model by varying the gravity harmonic coefficient perturbations directly with frequency dependence. 
We did not find any solution which could satisfactorily reproduce the data with g-modes.
We also explored the possibility of using a given mode's eigenfrequency as a tunable parameter, varying our expectation for each f-mode's eigenfrequency between half and thrice its theoretically predicted value. 
Although some frequencies fit better than others, none came close to the goodness of fit we obtain with our spectral model. 
We also tested the full pipeline omitting one flyby, testing each subset of four flybys to ensure the results were consistent, and not a spurious peculiarity of these five particular flybys. 
That is, we wanted to ensure that if one of the flybys had had a problem such that it did not successfully transmit data, that it would not have altered our conclusion. 
As expected, with fewer flybys the preferred region of parameter space could not be as tightly constrained, but the results were consistent and favored the same region shown in Figure~\ref{probs}. 
If future missions can perform the same experiment with a larger number of flybys, we may be able to make stronger conclusions. \\

All plausible spectral models predict large peak mode amplitudes on the order of several kilometers for a small number of modes (of order 5-10) near the peak frequency. 
Mode amplitudes inferred from velocity map time series of Jupiter are of order 100m (\cite{gaulme+2011}), so in order to explain our findings we require the peak amplitudes to be at least an order of magnitude larger on Saturn than have been observed on Jupiter. \\

\begin{figure}
\begin{subfigure}{.5\textwidth}
  \centering
  \includegraphics[width=\linewidth]{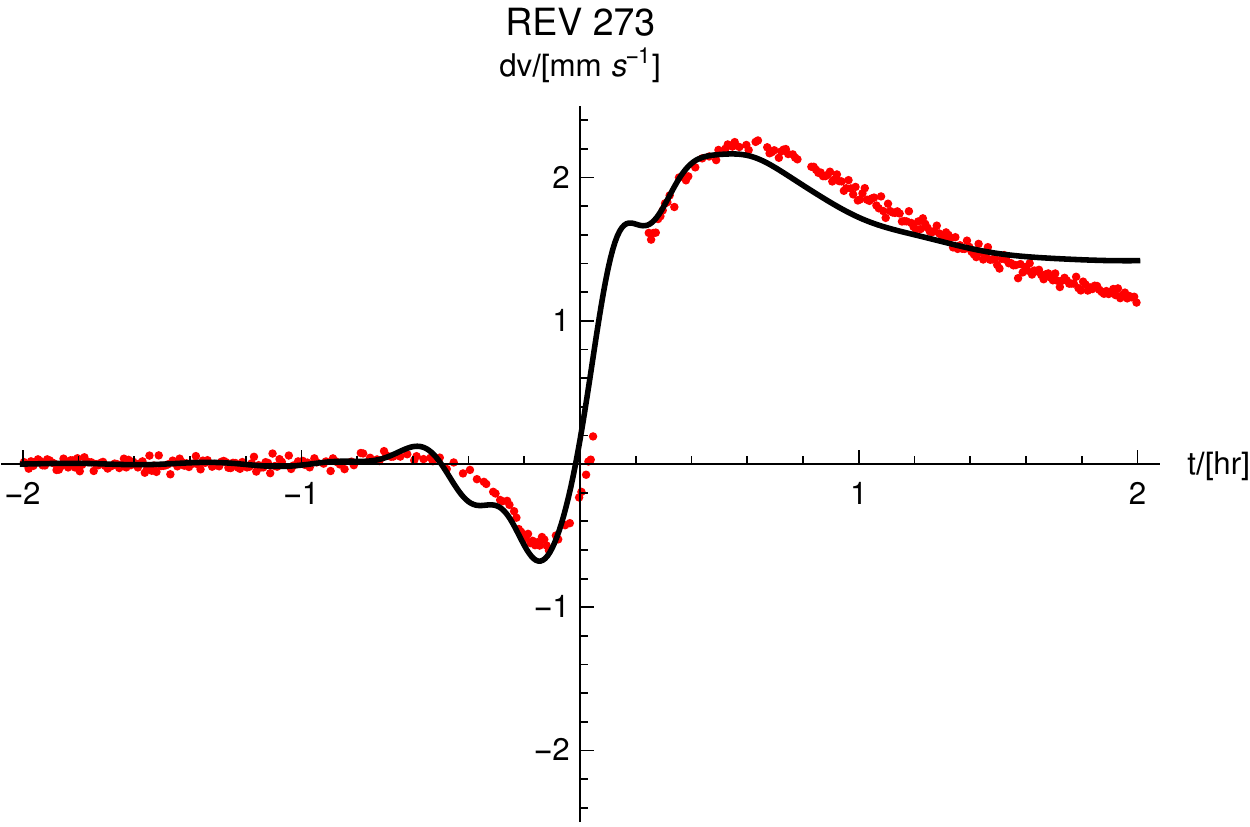}
  \label{fig:sfig1}
\end{subfigure}%
\begin{subfigure}{.5\textwidth}
  \centering
  \includegraphics[width=\linewidth]{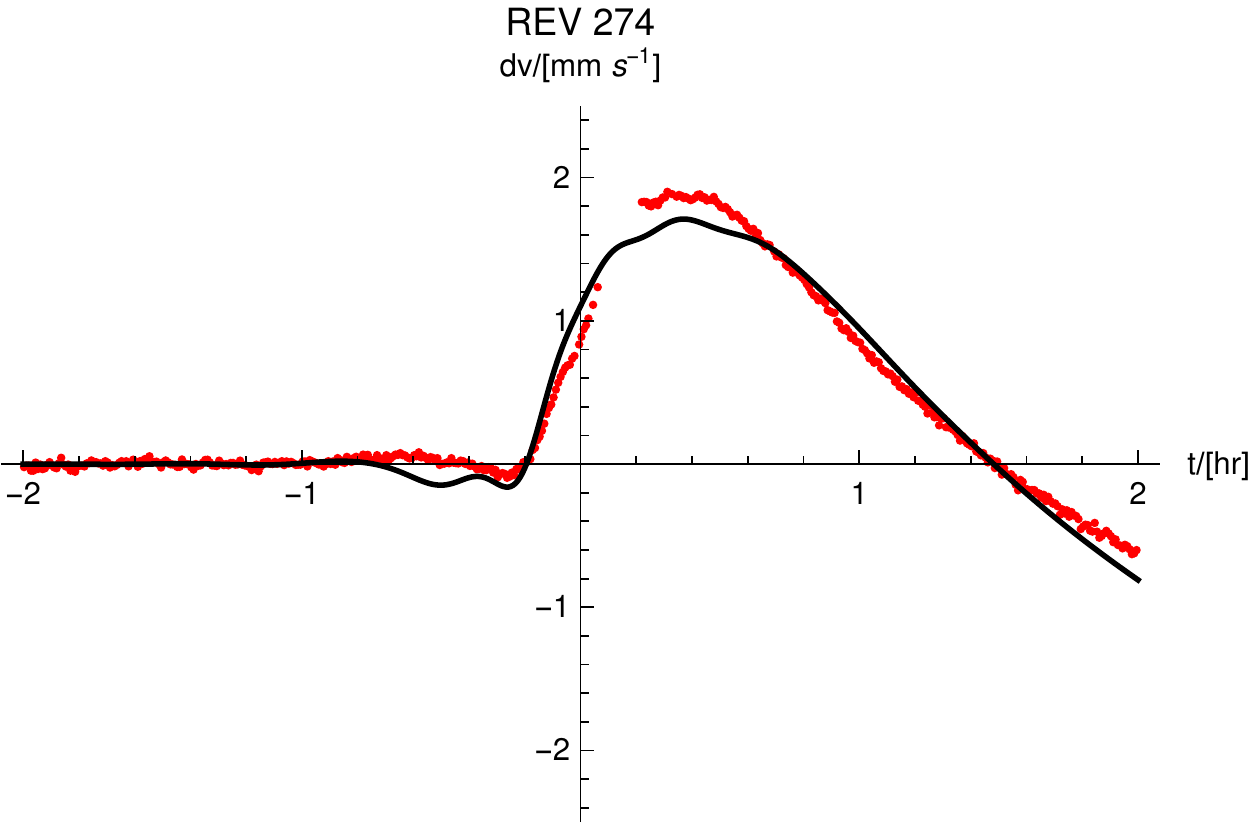}
  \label{fig:sfig2}
\end{subfigure}\\
\begin{subfigure}{.5\textwidth}
  \centering
  \includegraphics[width=\linewidth]{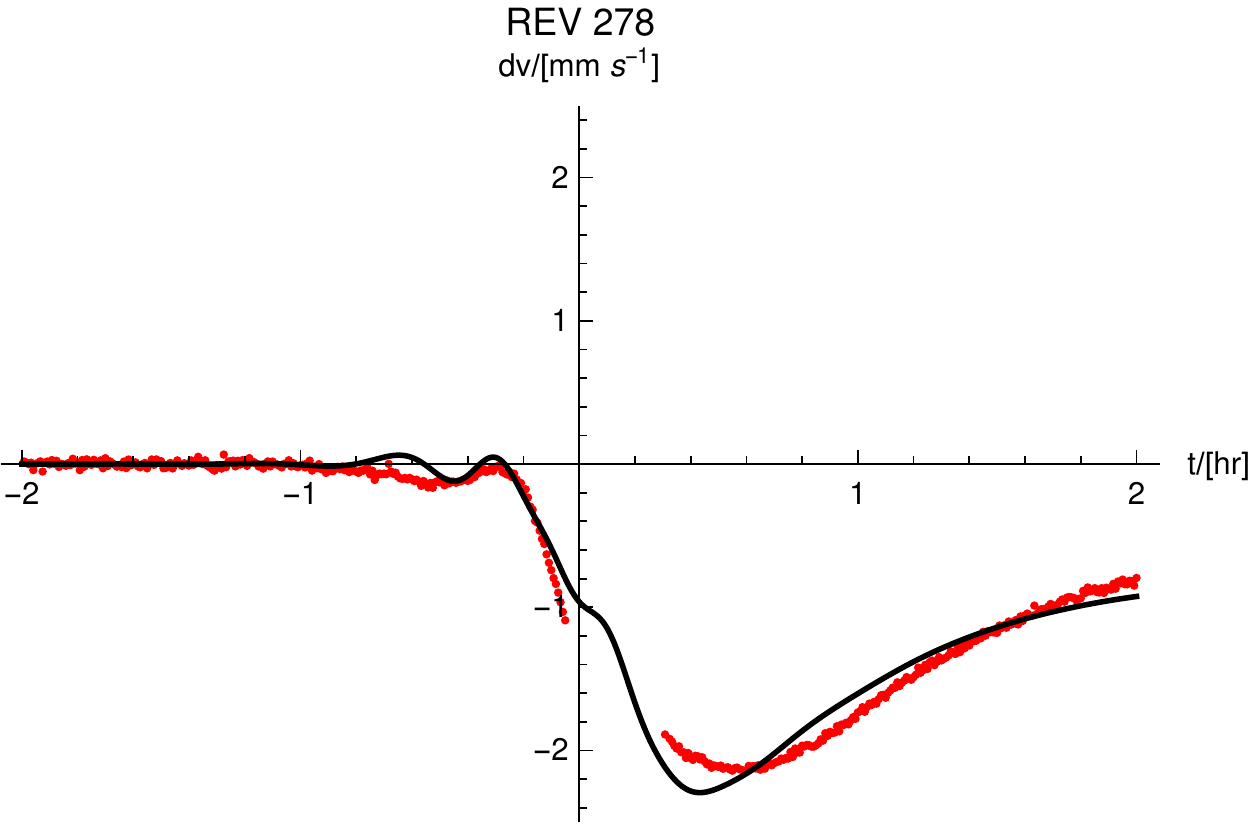}
  \label{fig:sfig1}
\end{subfigure}%
\begin{subfigure}{.5\textwidth}
  \centering
  \includegraphics[width=\linewidth]{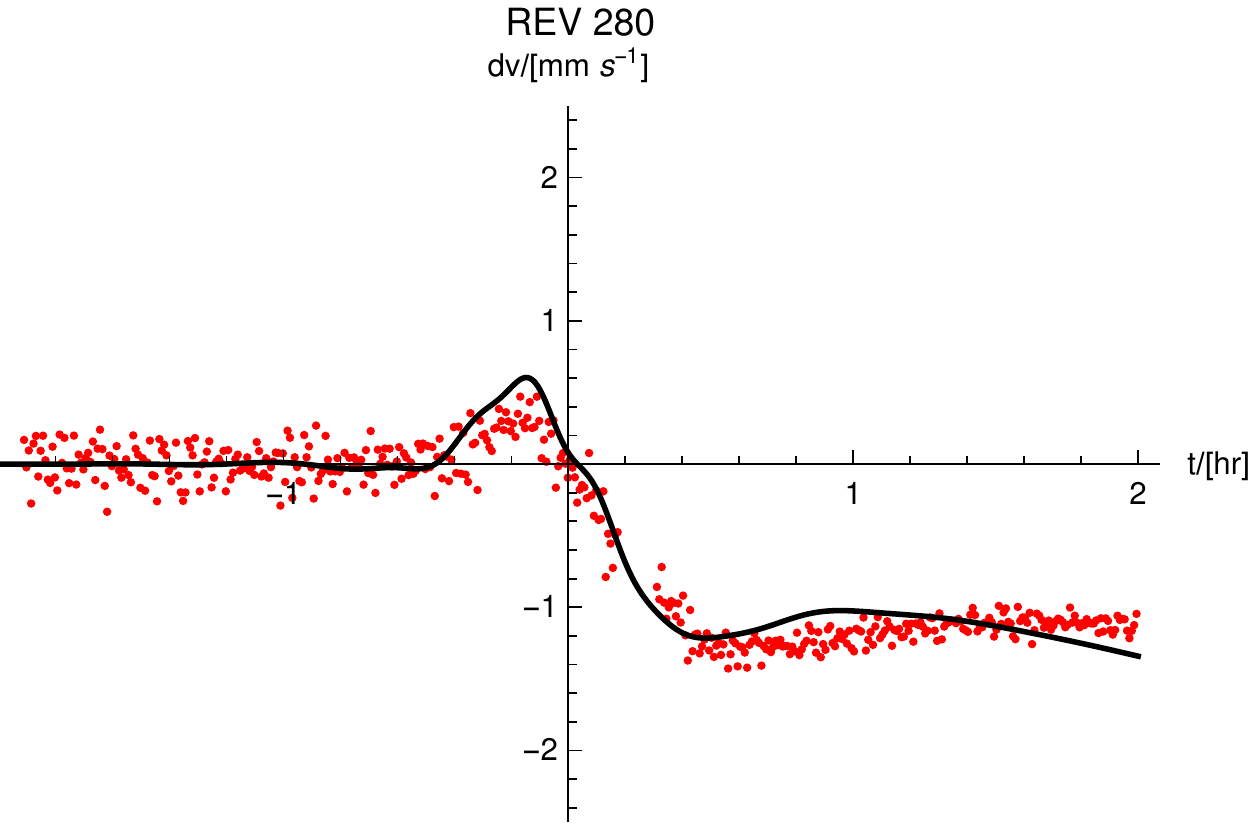}
  \label{fig:sfig1}
\end{subfigure}\\
\begin{subfigure}{.5\textwidth}
  \centering
  \includegraphics[width=\linewidth]{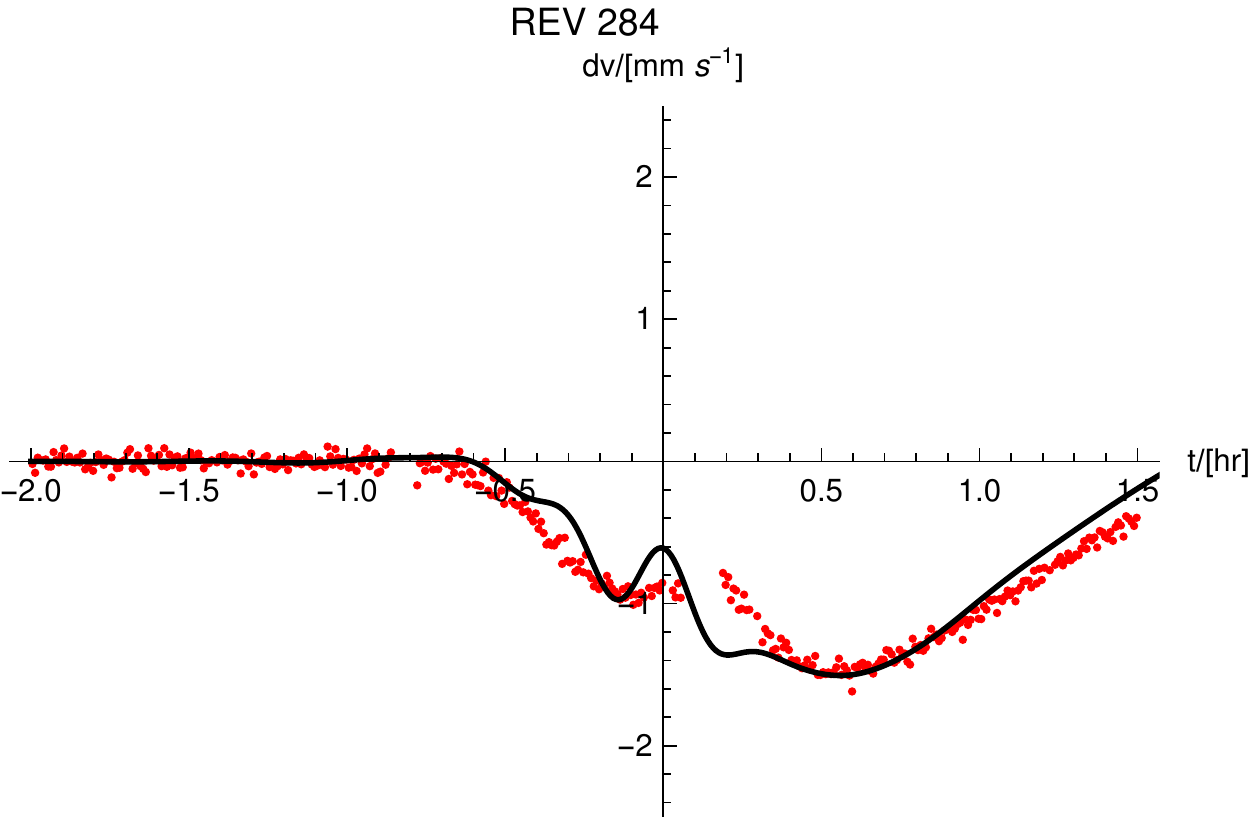}
  \label{fig:sfig1}
\end{subfigure}%
\caption{Forward calculation fits for individual flybys using our simple spectral model. The parameters used here are $\omega_0=600\mu$Hz and $\sigma=40\mu$Hz. The red scattered points are the data, and the black curves are the best fits models.}
\label{flyby-fits}
\end{figure}

This method also allows us to fit the range rate residuals from each individual flyby. 
We begin with a sample amplitude spectrum with a high probability of reproducing the data (see Figure~\ref{probs}). 
We then run a suite of simulations with random initial phases of each mode and show the best fit results for each flyby in Figure~\ref{flyby-fits}. 
Each fit optimizes for the best fit scaling coefficient, and all are in agreement within a factor of two. 

\section{Discussion}
We can reproduce the behavior of the non-zonal and/or non-static component of Saturn's gravity field using a simple three parameter forward model for mode amplitudes (Gaussian peak, width, and scaling). 
Our model uses a simple interior model and is not sensitive to detailed assumptions about Saturn's interior structure, spin rate, or rotation profile. 
For the amplitude spectrum of the modes, we rely only on a general understanding of Saturn's eigenfunctions and the equation of continuity in order to compute its gravitational effect for a given amplitude. 
We can also compute the mode inertia to scale the gravity signal for a given mode energy. \\

The best way to think of these results is in the Bayesian sense; we do not claim incontrovertible proof of p-modes on Saturn. 
Rather, given this particular data set, we present the probability that different models will reproduce this data in Figure~\ref{probs}. 
These probabilities should be used to update prior assumptions about Saturn's normal mode spectrum, bearing in mind that more complex models with more degrees of freedom (e.g. from Iess et al 2019) are not captured by our analysis. \\

We find a moderate preference for models which have a frequency peak between about 500 and 700 $\mu$Hz with a narrow width, although models with peaks as low as 250$\mu$Hz or as high as $1000\mu$Hz also have nonzero probability. 
Intriguingly, the inferred narrowness of the peak is analogous to the narrowly peaked five minute modes observed on the sun. 
Observing the power spectrum of the solar modes, one finds a peak frequency near 3000 $\mu$Hz with a full width half maximum (FWHM) of order 300 $\mu$Hz (\cite{frohlich+1997}), indicating a ratio of the FWHM to peak frequency of order 1/10.
Converting the standard deviation of our Gaussian frequency dependence functions to the corresponding FWHM, we likewise obtain a solution of order 1/10 (for example in the case of the fit shown in Figure~\ref{example}.)
This may indicate some similarity between the two systems, for example that the peak frequency is set by some dynamical process with a characteristic timescale. 
On the sun this timescale is the eddy turnover time in the top scale height of the convective zone. 
Although the same mechanism cannot excite the observed amplitudes on Saturn, less frequent moist convective events with a characteristic turnover timescale (see e.g. \cite{markham-stevenson2018}) could produce similar strong frequency dependence.  
On Jupiter, new theories to explain the ammonia distribution require updrafts which traverse 100km in 1000s (\cite{guillot+2020}). 
If similar dynamics occur on Saturn, the timescale is roughly consistent with the peak frequencies inferred by this work. 
Others have suggested a large impact as a source of Saturn's oscillations (\cite{wu-lithwick2019}). 
Although a 150km impactor could in principle excite km-scale oscillations in p-modes, the scaling suggests the gravity signal from f-modes should always dominate. 
Therefore if this is indeed the dominant excitation mechanism on Saturn, there must be some other reason to preferentially dissipate f-modes or preferentially amplify p-modes.
\\

Most notably, we cannot reproduce the time series data with f-modes; neither with a single f-mode dominating the signal nor with a straightforward superposition of f-modes. 
This finding is consistent with inferred amplitudes of f-modes which have been measured using Saturn's rings (kronoseismology) (\cite{hedman-nicholson2013}, \cite{wu-lithwick2019},\cite{fuller2014}), which are determined to be on the order of a meter in amplitude and should not produce this large of a signal in Cassini's gravity experiment. 
Assuming the amplitudes inferred from ring data, the detectability should have been marginal ($\delta v \sim 0.05$mm~s$^{-1}$ instead of the observed $\sim 2$mm~s$^{-1}$). 
The p-modes that are required to produce the observed signal would need surface amplitude on the order of kilometers, implying radial velocities of meters per second. 
These p-modes despite their large amplitudes, are not expected to show structure in the rings, because the relevant resonant radius for these frequencies is well inside Saturn's C ring. \\

The required p-mode amplitudes are at least an order of magnitude larger than were observed on Jupiter (\cite{gaulme+2011}). 
Interestingly, early analysis of the Juno mission indicates a similar unexplained gravity signal on Jupiter that is approximately 20 times weaker than the signal observed on Saturn (\cite{durante+2020}). 
This is interesting, because this analysis indicates that if the relevant amplitudes were those observed by Gaulme et al 2011, then we should expect a similar time dependent signal diminished in scale by about an order of magnitude. 
Replicating our analysis of Cassini's gravity data for Juno, which has many more planned gravity orbits than Cassini, may be a promising future application of the method outlined in this paper. 
We can test to see if the inferred normal mode spectrum from gravity measurements on Jupiter is consistent with the corresponding power spectrum obtained with Earth-based observations. \\

Because we predict large peak amplitudes, we must consider if these are plausible and consistent with existing data. 
Voyager radio occultation measurements of Saturn has error estimates between 6 and 10km, and the measurements found incompatible radii between the northern and Southern hemisphere on the order of 10km (\cite{lindal+1985}). 
These uncertainties are compatible with the time dependent shape variations our analysis predicts. 
Our analysis here would predict future measurements of Saturn's shape cannot obtain better accuracy than around a few kilometers. 
A series of highly accurate measurements of Saturn's shape should have a time dependent component on the order kilometers. \\

We must also consider how our findings can constrain $Q$. 
Using a set of $N$ modes excited to 10km amplitude $a$ (a high estimate, see Table~\ref{tab:table1}), powered by the full luminosity of Saturn $\mathcal{L}$ as a lower bound on $Q$, we compute 
\begin{equation}
Q > \frac{a^2 \omega^3 M N}{\mathcal{L}} \sim 10^7
\end{equation}
 for the relevant peak modes we have identified, where $M$ is the modal mass and $N\sim 5-10$ corresponds to the number of modes with significant amplitude (for example, greater than a kilometer). 
This $Q$ is compatible with estimates so far based on theory (\cite{markham-stevenson2018}, \cite{wu-lithwick2019}). \\

We must also check that the modes can still be approximated as linear perturbations, i.e. $\mathbf{u} \cdot \nabla \mathbf{u} \ll \frac{d \mathbf{u}}{dt}$ where $\mathbf{u}$ is the velocity vector. 
For the peak modes we identified, the frequency is sufficiently high that the motion is almost purely vertical. 
$\frac{du}{dz}$ is most significant near the surface when the atmospheric properties vary quickly. 
We know in this region $\frac{du}{dz} \sim \left(1-\sqrt{1-\left(\frac{\omega}{\omega_c}\right)^2}\right) u/H$ (\cite{christensen-dalsgaard}), where $H$ is the scale height and $\omega_c$ is the acoustic cutoff frequency. 
Therefore the condition that the system can be treated linearly is 
\begin{equation}
\left(1-\sqrt{1-\left(\frac{\omega}{\omega_c}\right)^2}\right) \ll \omega H/u
\end{equation}
the ratio of the left hand side to the right hand side using $\omega \sim 5\times 10^{-3}$ for 1km amplitude modes is about $10^{-3}$. 
This is the maximum value near the surface; the value is much smaller in the interior where most of the mode inertia is. 
If the inferred amplitudes are correct, nonlinear effects are important for p-modes on both Saturn and Jupiter. \\

Many questions remain, and it is clear that the field of giant planet seismology--both observational and theoretical--is in its infancy. 
Here we demonstrate that the unexpected and unexplained components of Saturn's ``dark'' gravity field can be straightforwardly modeled as simple frequency dependent seismic activity. 
This provides one more piece of plausible evidence that the giant planets are seismically active, and should motivate further observations and theoretical study.

\bibliography{library}{}
\bibliographystyle{apalike}

\section{Appendix}
\begin{table}[h!]
\begin{center}
\caption{Sample spectrum using $\omega_0=600mu$Hz and $\sigma=40\mu$Hz, listing modes with amplitudes larger than 100m. This spectrum should not be taken too seriously as good fit solutions are degenerate and non-unique--the general orders of magnitude should be paid attention to more than the specific modes.}
\label{tab:table1}
\begin{tabular}{l|c|r}
\textbf{(n,l)} & $\omega_{\text{nlm}}/[\mu$Hz] & $a_{\text{nlm}}/$[m] \\
\hline
(2,4) & 497 & 107\\
(2,5) & 525 & 461\\
(2,6) & 551 & 1220\\
(2,7) & 575 & 2140\\
(2,8) & 597 & 2700\\
(3,2) & 553 & 1190\\
(3,3) & 599 & 2350\\
(3,4) & 633 & 1900\\
(3,5) & 662 & 965\\
(3,6) & 689 & 334\\
(4,2) & 667 & 794
\end{tabular}
\end{center}
\end{table}

\subsection{Potential perturbation}
In the following derivation, we will assume the spacecraft is on a prescribed Keplerian orbit, and assess the gravity potential field it encounters as a function of time. 
In reality, the data we have is a velocity time series; we cannot measure the gravity potential directly. 
However, we demonstrate how a gravity potential with stochastic elements can be averaged toward deterministic behavior that isolates information about the amplitude spectrum. 
This same basic procedure will be employed to forward model the velocity time series, although the details in that case are considerably more complicated. 
\\

We begin with Equation~\ref{mode-pot}. 
The full potential of all modes for a given flyby $i$ is 
\begin{equation}
\Phi_i(t) = \sum_{\text{nlm}}A_{\text{nlm}} \left( \frac{R}{r(t)} \right)^{l+1} P_{lm}(\cos(\theta(t))) \cos[m(\phi(t)-\Omega t-\phi_i)-\omega_{\text{nlm}}t-\alpha_{\text{nlm,i}}] 
\end{equation}
with $A_{\text{nlm}}\equiv a_{\text{nlm}}\delta C_{\text{nlm}}$. $\phi_i$ is the random initial longitudinal orientation of Saturn (the same for each mode, but random for each orbit because Saturn's spin rate is not precisely known), and $\alpha_{\text{nlm,i}}$ is the temporal phase of each mode--random for each mode and for each flyby because we do not know the eigenfrequency with sufficient precision to impose phase coherency between subsequent close encounters. 
The uncertainty in $\phi_0$ can be absorbed into $\alpha$; for zonal modes, $\phi_0$ does not matter, and for tesseral/sectoral modes it can be added into $\alpha$ so that there is only one relevant random variable. 
We can rewrite this expression using the harmonic addition theorem:
\begin{dmath}
\Phi_i(t) = \sum_{\text{nlm}} A_{\text{nlm}} f_{\text{lm}}(t) \left(\cos(g_m(t))\cos\alpha_{\text{nlm,i}} - \sin(g_m(t)) \sin\alpha_{\text{nlm,i}} \right)
\end{dmath}
where $f_{\text{lm}}(t)=\left( \frac{R}{r(t)} \right)^{l+1} P_{lm}(\cos(\theta(t)))$ and $g_m(t)=m(\phi(t)-\Omega t)-\omega t$ is the same for every orbit and does not depend on the random variable $\alpha_{\text{nlm,i}}$. 
Now we square this expression 
\begin{dmath}
\Phi_i(t)^2 = \sum_{\text{q}} \sum_{\text{q'}} A_{\text{q}} A_{\text{q'}} f_{\text{q}}(t) f_{\text{q'}}(t) (\cos(g_q(t)) \cos(g_{q'}(t)) \cos \alpha_{\text{q,i}} \cos\alpha_{\text{q',i}} -\cos(g_q(t))\sin(g_{q'}(t)) \cos\alpha_{\text{q,i}}\sin\alpha_{\text{q',i}} - \sin(g_q(t))\cos(g_{q'}(t)) \sin\alpha_{\text{q,i}} \cos\alpha_{\text{q',i}} + \sin(g_q(t))\sin(g_{q'}(t)) \sin\alpha_{\text{q,i}} \sin\alpha_{\text{q',i}}) 
\end{dmath}
where we substitute a single index $q$ to refer to a given mode $(n,l,m)$ for notation convenience. \\

From here, we perform the crucial step of summing over many such flybys. 
This total can be expressed
\begin{dmath}
\sum_{i=1}^N \Phi_i(t)^2 = \sum_{i=1}^N \sum_{\text{q}} \sum_{\text{q'}} A_{\text{q}} A_{\text{q'}} f_{\text{q}}(t) f_{\text{q'}}(t) (\cos(g_q(t)) \cos(g_{q'}(t)) \cos \alpha_{\text{q,i}} \cos\alpha_{\text{q',i}} -\cos(g_q(t))\sin(g_{q'}(t)) \cos\alpha_{\text{q,i}}\sin\alpha_{\text{q',i}} - \sin(g_q(t))\cos(g_{q'}(t)) \sin\alpha_{\text{q,i}} \cos\alpha_{\text{q',i}} + \sin(g_q(t))\sin(g_{q'}(t)) \sin\alpha_{\text{q,i}} \sin\alpha_{\text{q',i}}) 
\end{dmath}
In this case, we can assume random variables behave as true statistical averages. 
We make use of the fact that $\sum_{q,q',i,j}\cos \alpha_{\text{q,i}} \cos \alpha_{\text{q',j}} \rightarrow \delta_{qq'}\delta{ij}$. 
Similarly $\sum_{q,q',i,j}\cos \alpha_{\text{q,i}} \sin \alpha_{\text{q',j}} \rightarrow 0$. 
Using this asymptotic behavior, we can write for a sufficiently large number of $N$ flybys,
\begin{equation}
\sum_i^N \Phi_i^2(t) \sim \frac{N}{2} \sum_{\text{nlm}} A_{\text{nlm}}^2 f_{\text{nlm}}^2(t)
\end{equation}
Thus a particle on a prescribed trajectory encountering potential perturbations due to normal modes will, when averaging over many such encounters, approach a deterministic curve that does not depend on the initial phase of each mode.
\\

\subsection{Acceleration and velocity}
The gravity experiment does not directly measure the gravity potential. 
The data we have is the velocity perturbation along a single axis. 
Our data stacking method is most straightforwardly derived for gravity potential perturbations.
To calculate gravitational acceleration we apply $\delta g_{nlm} = - \nabla \delta \Phi_{nlm}$ to obtain 
\begin{dmath}
\frac{\delta \mathbf{g}_{nlm}}{A_{nlm}} = \frac{G M}{r^2} \left(\frac{R}{r}\right)^l \left[ 
-(l+1) \cos[m(\phi-\Omega t)-\omega t-\alpha]P_l^m(\cos\theta) \hat{\mathbf{r}} 
+ \cos[m(\phi-\Omega t)-\omega t-\alpha] \frac{\partial P_l^m(\cos\theta)}{\partial \theta} \hat{\theta} 
- \frac{m}{\sin\theta} \sin[m(\phi-\Omega t)-\omega t-\alpha] P_l^m(\cos\theta) \hat{\phi} \right]
\end{dmath}
This gives us the components we need to project onto $\hat{\oplus}$, the unit vector pointing toward Earth. 
The total gravitational acceleration perturbation from normal modes can be written 
\be 
\delta g_\oplus(t) = \sum_{nlm} A_{nlm} g_{\oplus,nlm}(t)
\ee
where we can express 
\be 
g_{\oplus,nlm}(t) = A_{nlm} \left( f_1(t) + f_2(t) \right)(f_3(t)\cos\alpha + f_4(t)\sin\alpha)
\ee
where $f_i(t)$ are tedious but nevertheless well defined functions of time independent of $\alpha$. 
Just as above, if we have an expression in this form, we can express the asymptotic average of the sum of a large number of flybys as 
\be 
\sum_{i=1}^N g_{\oplus,i}^2(t) = \frac{N}{2} \sum_{nlm} A_{nlm}^2 f_{nlm}(t)^2
\ee
where $f(t)^2$ is a function of $f_1(t)$ and $f_2(t)$. \\

A similar procedure can be followed for velocity perturbations, and this is the source of frequency dependence (absent in averaged potential perturbations) $\omega_{\text{nlm}}$ by temporally integrating the gravitational perturbations.
In fact it is not this simple; a perturbed spacecraft will deviate from its Keplerian trajectory, that leads to errors from the exact solution of order $(\Delta r/r)\Delta t$. 
A better approximation is to dynamically solve the equation of motion, accounting for displacement from the initial Keplerian trajectory to linear order, although this method produces higher order errors from the exact solution. 
The method we eventually used, solving the exact equation of motion explicitly then subtracting the Keplerian solution, is not conducive to an analytic expression. 
Because of this, we verified these results by testing them numerically against simulations to be sure they did approach a deterministic curve when stacked. \\

To do this, we performed $10^4$ Monte Carlo simulations of random flybys for each mode. 
We then averaged random subsets of these samples to verify that they asymptotically approach an asymptotic curve. 
We also directly simulated various examples of a spectral superposition of different modes, running $10^4$ flybys. 
We then compared these stacked results against the superposition of a spectrum of averaged squared modes, finding excellent agreement with Equation~\ref{velocity-stack}. 
We therefore have an analytic approximation which motivates the stacking procedure, as well as more exact numerical tests to make sure the results from the analytic approximation are robust to this application. 
These tests verify the validity of our data stacking procedure.
In practice, we have a finite number $N=5$ flybys. 
A random set of 5 flybys will deviate somewhat from the asymptotic behavior of $10^4$ flybys, so for the actual probabilistic fitting routine we used Equation~\ref{velocity-exact}. 
We used the statistical method outlined in the Analysis and Results section to account for this, by directly solving for the probability that a given spectrum will produce a good fit to the data. \\

\subsection{Spacecraft sensitivity and model intuition}
\begin{figure}
\centering
\includegraphics[scale=1]{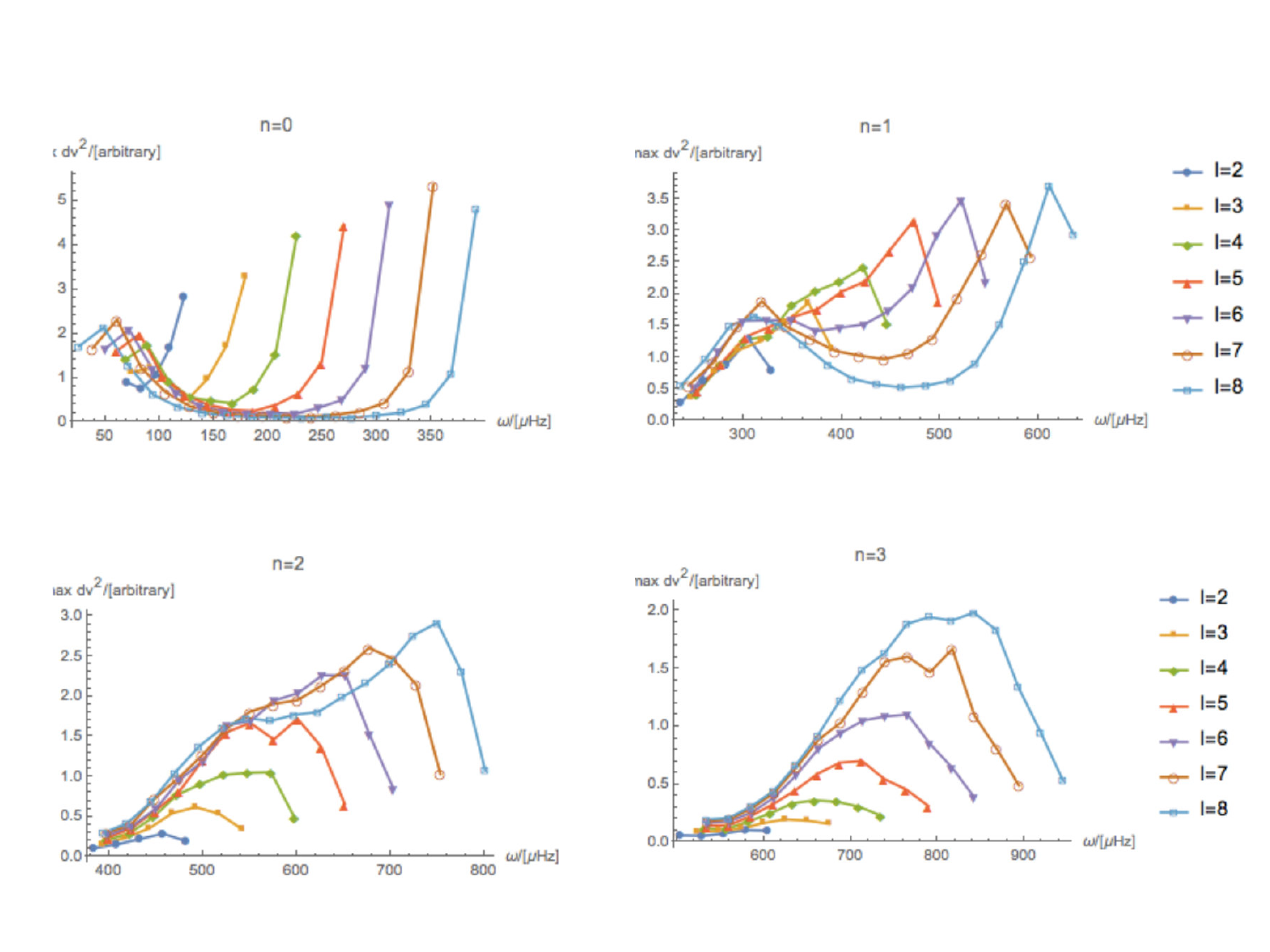}
\caption{In the above plots the x-axis is the frequency in the inertial frame in which Cassini's orbit is defined. This is why different m-values are so spread out in frequency when in Saturn’s rotating frame they are very close. The y-axis is the average maximum squared velocity response to a particular mode, using fixed gravity potential perturbation coefficients for each mode. }
\label{intuition}
\end{figure}

We must consider whether the particular orbit of Cassini biases its ability to detect certain modes. 
If the convolution of a particular mode's space dependent eigenfunction and time dependent eigenfrequency appears stationary in Cassini's frame during close approach, then Cassini will preferentially detect signals from this mode. 
For example, because Cassini's orbit is nearly polar, there may be an intrinsic preference for sectoral ($m=\pm l$) modes, as seems to be true for f-modes (see Figure~\ref{intuition}). 
As a different example, if a mode's half period is near the timescale of Cassini's motion from the northern to the southern hemisphere, which appears to be the case for tesseral modes with $m= l - 1$ in Figure~\ref{intuition}. 
Moreover, because Cassini's orbit moves from west to east, it is plausible that it would preferentially detect $m>0$ modes whose pattern rotates in the prograde direction, an intuition also supported by Figure~\ref{intuition}. 
According to Figure~\ref{intuition}, this quasi-resonant effect does indeed make Cassini more sensitive to certain modes. 
All these effects are implicitly accounted for in the forward model in the main text.
\\

Another important consideration is high degree f-modes. 
If the strong frequency dependence we predict applies equally to f-modes, then we must consider if Cassini would detect them. 
There are two sources of attenuation: the intrinsic geometric attenuation with distance of high degree gravity harmonics $(R/r)^l$, and the monotonically decreasing gravity potential coefficient response to each mode which obeys $(l(l+1))^{-1/2}$ (\cite{wu-lithwick2019}). 
We use a characteristic radius for Cassini's close encounter $r_{c} \sim \frac{v}{R} \int_{-R/2v}^{R/2v} r(t) dt \sim 1.15 R$ where $v$ is the periapse velocity. 
High degree f-mode eigenfrequencies obey $\omega^2 \sim \frac{GM}{R^3}(l(l+1))^{1/2}$, so the modes in the most likely peak frequency region have $l \sim 50-100$. 
Using our attenuation estimates, these modes would be reduced by at least 5 orders of magnitude compared to low order f-modes, significantly more severe than the 2-3 orders of magnitude reduction in low order p-modes. 
Therefore, we do not expect for Cassini to detect high order f-modes, even if they likewise had km-scale amplitudes. 

\end{document}